\documentclass[%
 reprint,
 amsmath,amssymb,
 aps,
]{revtex4-2}

\usepackage{graphicx}
\usepackage{dcolumn}
\usepackage{bm}
\usepackage{bbm}
\usepackage{amsmath,amsthm,amssymb,bm}
\usepackage{physics}
\usepackage{old-arrows}
\usepackage{leftindex}
\usepackage{tikz}
\usetikzlibrary{shapes}
\usepackage[normalem]{ulem} 
\usepackage{xcolor}
\definecolor{customblue}{HTML}{D6E8F5}

\makeatletter
\newdimen\@myBoxHeight%
\newdimen\@myBoxDepth%
\newdimen\@myBoxWidth%
\newdimen\@myBoxSize%
\newcommand{\SquareBox}[2][]{%
    \settoheight{\@myBoxHeight}{#2}
    \settodepth{\@myBoxDepth}{#2}
    \settowidth{\@myBoxWidth}{#2}
    \pgfmathsetlength{\@myBoxSize}{max(\@myBoxWidth,(\@myBoxHeight+\@myBoxDepth))}%
    \tikz \node [shape=rectangle, shape aspect=1,draw=red,inner sep=2\pgflinewidth, minimum size=\@myBoxSize,#1] {#2};%
}%
\makeatother

\begin{document}

\preprint{Published in Phys. Rev. A 112, 052402 (2025).
DOI: \href{https://doi.org/10.1103/vcqj-xd7z}{10.1103/vcqj-xd7z}}

\title{Exact quantification of bipartite entanglement\\in unresolvable spin ensembles}

\author{Tzu-Wei Kuo}
 \email{tka71@sfu.ca}
\author{Hoi-Kwan Lau}%
 \email{hklau.physics@gmail.com}
\affiliation{Department of Physics, Simon Fraser University, Burnaby, British Columbia, Canada V5A 1S6}%

\date{\today}

\begin{abstract}
Quantifying mixed-state entanglement in many-body systems has been a formidable task. In this work, we quantify the entanglement of states in unresolvable spin ensembles, which are inherently mixed. By exploiting their permutationally invariant properties, we show that the bipartite entanglement of a wide range of unresolvable ensemble states can be calculated exactly. Our formalism is versatile; it can be used to evaluate the entanglement in an ensemble with an arbitrary number of particles, effective angular momentum, and bipartition. We apply our method to explore the characteristics of entanglement in different physically motivated scenarios, including states with definite magnetization and metrologically useful superpositions such as Greenberger-Horne-Zeilinger (GHZ) states and spin-squeezed states. Our method can help understand the role of entanglement in spin-ensemble-based quantum technologies. 
\end{abstract}

\maketitle


\section{\label{Introduction}Introduction}
Solid-state or atomic spin ensembles, such as diamond NV centers \cite{NV_center1, NV_center2, NV_center3}, semiconductor quantum dots \cite{quantum_dots1, quantum_dots2}, ions in Penning traps \cite{Ion_trap_1, Ion_trap_2}, or atomic clouds \cite{atomic_clouds1, atomic_clouds2}, are promising platforms for implementing quantum technologies \cite{atom_number, Quantum_interface, microwave_photons, single-photon_source, telecom, time-bin}. Because electronic and nuclear spins usually interact weakly with their environment and among themselves, spin ensembles generally exhibit long-lived coherence times, from seconds for electronic spins \cite{one_second_electronic_spins} up to hours for nuclear spins \cite{hours_nuclear_spins, Simmons}. Spin ensembles also play an integral role in many hybrid architectures \cite{scalability, hybrid1, hybrid2} due to their collectively enhanced coupling strength \cite{superconducting1, superconducting2}. These properties make spin ensembles favorable platforms for implementing quantum sensors \cite{sensing, quantum_metrology}, quantum repeaters \cite{repeater}, quantum computers \cite{atom_number, quantum_dots3}, and as testbeds for simulating light-matter interactions \cite{superradiance, subradiance}.

Many ensemble-based technologies rely on the entanglement among spins to realize quantum advantages. For example, in sensing applications, the sensitivity of an unentangled probe is bounded by the so-called standard quantum limit, but this can be surpassed by using highly entangled states such as the Greenberger-Horne-Zeilinger (GHZ) states or spin-squeezed states \cite{Metrology1}. Moreover, entangled quantum emitters coupling to the same light field can exhibit superradiance \cite{superradiance1}, which can be used to make stronger and more stable lasers \cite{superradiance2}. Furthermore, the decoherence-free subspace for protecting quantum information generally requires logical information to be encoded in many-body entangled states \cite{DFS}. To understand the role of entanglement in these applications, quantifying the amount of entanglement in many-body resource states is essential.

In addition, from a computational perspective, the amount of entanglement usually determines whether a many-body system can be efficiently simulated by classical means. In fact, classical algorithms, such as the density-matrix renormalization group (DMRG) \cite{DMRG_method} and tensor network methods, could fail if entanglement grows too fast with the size of the system \cite{Volume_law_1, Volume_law_2}. Therefore, quantifying entanglement in a many-body state would help identify tractable problems and potentially justify the use in cases of quantum computing.

Entanglement quantification has other technological applications. For example, sensitivity has been suggested as a witness of many-body entanglement \cite{app_benchmark1}, so an exact quantifier can be used as a benchmark. Also, since entanglement is a resource for many quantum technologies, quantifying entanglement can evaluate the quality of the gates and systems \cite{app_benchmark2}.

However, quantifying entanglement in spin ensembles is inherently challenging due to two main factors. First, the dimension of the Hilbert space $\mathcal{H}_N$ grows exponentially with the number of particles $N$. Brute-force quantification methods will quickly become infeasible even for ensembles with just tens of spins. Second, many ensemble states of interest are inherently mixed. Explicitly, most ensemble-based platforms suffer from limited resolvability of individual spins, because they are usually manipulated by an auxiliary system whose physical size is much larger than the spin-spin separation \cite{unresolvability}, or the spins are rapidly moving so that tracking is technologically challenging \cite{Rapid_moving_1, Rapid_moving_2}. Without spin resolvability, we can access only the collective but not individual properties of the spins; missing the complete information means that the description of the ensemble state is inherently mixed \cite{PI_definition1, Nori}. Unfortunately, exactly quantifying entanglement in mixed states requires optimizing over all possible state decompositions, which is known to be a formidable task \cite{Characterizing_entanglement_for_symmetric_systems, Entanglement_measures}.

Few strategies exist for computing entanglement for mixed-state many-body systems. Analytical expressions are known only in limited cases, usually in low dimensions \cite{Entanglement_measures}. Therefore, most common approaches involve numerical simulations using semidefinite programming \cite{Semi1, Semi2, Semi3} or tensor network techniques \cite{Tree_tensor, Tensor_network_sampling}. However, these methods are often computationally expensive and may not provide much intuition about the underlying physics. Alternatively, many studies in the literature consider entanglement witnesses that fit the experimental configuration or the mathematical structure of the spin ensemble states \cite{Entanglement_witness_1, Entanglement_witness_2}. In particular, witness-based approaches have been developed to place experimentally accessible bounds on spin-ensemble entanglement, from general frameworks that link witnesses to entanglement quantifiers \cite{entanglement_quantification_using_witnesses} to more recent techniques tailored to atomic ensembles \cite{entanglement_quantification_in_atomic_ensembles}. While powerful, these approaches can only bound, not exactly quantify, the amount of entanglement. Such bounds are often too coarse to reveal the detailed scaling of entanglement with system parameters, limiting their usefulness for benchmarking resource states or optimizing applications. Moreover, they typically provide lower bounds to measures such as the best separable approximation or the generalized robustness, whose interpretations as robustness to noise or distance from separability are not as directly tied to operational tasks. In contrast, the entanglement of formation is directly associated with entanglement cost or resource conversion, making it a more natural quantity to target for entanglement quantification.

\begin{figure*}[htp!]
    \centering
    \includegraphics[width=16cm]{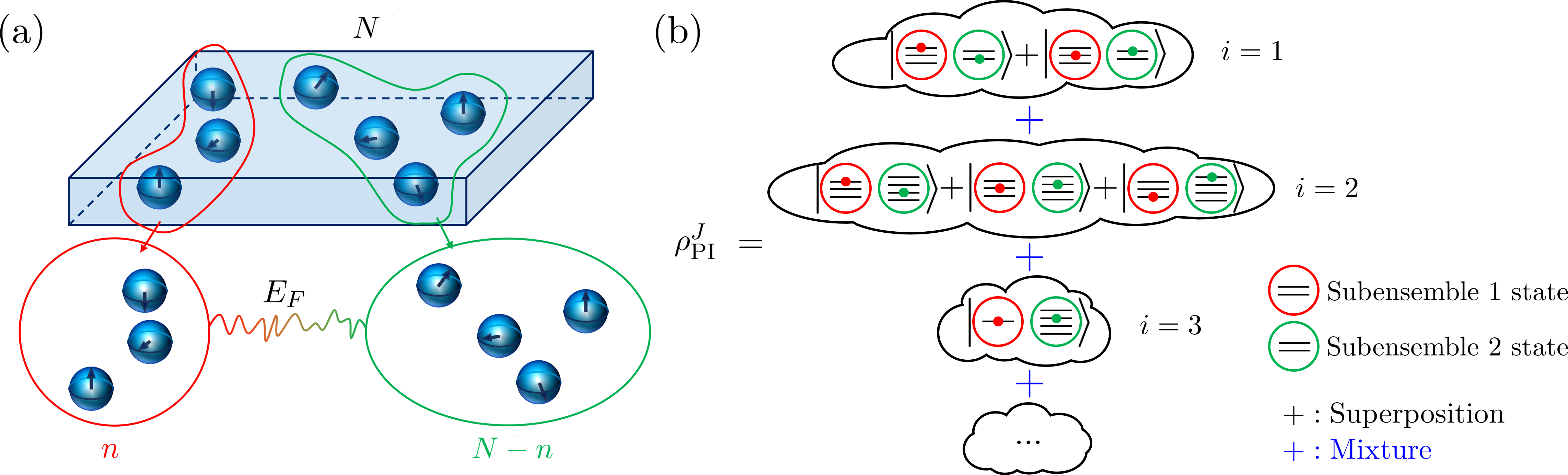}
    \caption{(a) An ensemble of $N$ spin-1/2 particles is divided into two subensembles: one with $n$ particles (red) and the other with $N-n$ particles (green). Without the ability to resolve individual spins, the ensemble state is permutationally invariant and generally mixed. Our goal is to exactly quantify the entanglement of formation $E_F$ between the subensembles, provided that all collective properties of the ensemble are known, i.e., logically pure. (b) A logically pure PI state $\rho_{\text{PI}}^J$ is generally an equal mixture of all degenerate states that share the same collective properties, cf. Eq.~(\ref{Superposition PI state}). Each degenerate component (denoted as cloud) is composed of subensemble states that belong to a different combination of subensemble angular momentum ($j_1$,$j_2$) and degeneracy indices ($i^{j_1}$,$i^{j_2}$). Since the subensemble bases are different, the superposition amplitudes and hence entanglement are generally different in each component. We show that the entanglement of formation of $\rho_{\text{PI}}^J$ is exactly given by the average entanglement entropy of all components in the standard decomposition, cf. Eq.~(\ref{Basis state being split with i}).}
    \label{Setup}
\end{figure*}

In this work, we provide a systematic way to exactly quantify the entanglement of formation of a broad class of mixed states with great practical relevance. Specifically, we consider permutationally invariant (PI) states, which is the general description of the states of unresolvable spin ensembles \cite{PI_definition1, Nori}. PI states are generally mixed even if we have complete information about their collective properties. For these states, we prove that their standard decomposition is also that which minimizes the average entanglement entropy, so their entanglement of formation $E_F$ can be computed exactly as the average entanglement of the composition states. Using the exact formalism, we explore how entanglement depends on ensemble parameters, such as the total angular momentum $J$, magnetization $M$, number of particles $N$, and partition number $n$, as illustrated in Fig.~\ref{Setup}a. For angular momentum eigenstates, we find that the entanglement behavior is generally different from Dicke states, which is the only class of PI state that is pure and thus well studied. Moreover, we extend our studies to metrologically relevant superposition states, such as GHZ-like states and spin-squeezed states. We note that the PI states studied in this work are different from the ``single-sided permutational symmetric" states in \cite{Characterizing_entanglement_for_symmetric_systems}, which impose more stringent and artificial restrictions on the density matrix of the state.

This paper is organized as follows. In Sec.~\ref{Section: Permutationally invariant states}, we review the background of PI states and discuss how, in principle, their decomposition can be resolved without demolition via local subensemble measurements. Leveraging this property, in Sec.~\ref{Section: Exact quantification of entanglement for PI states}, we prove that entanglement of formation of PI states can be exactly quantified and provide a closed-form expression. In Sec.~\ref{Section: Angular momentum eigenstates}, we show how the entanglement of an angular momentum eigenstate depends on various physical parameters. In Sec.~\ref{Section: Superposition states}, we quantify the entanglement of two classes of metrologically relevant superposition states: GHZ-like states and spin-squeezed states. We conclude our findings in Sec.~\ref{Section: conclusion}.

\section{\label{Section: Permutationally invariant states}Permutationally invariant states}
\subsection{\label{Subsection: Definition}Definition}
The state of an ensemble consisting of $N$ spin-1/2 particles can be expressed as a linear combination of the degenerate basis states $\{\ket{J,M,N,i}\}$ \cite{PI_definition1, Dicke_paper}. Every $\ket{J,M,N,i}$ state is a simultaneous eigenstate of the total angular momentum operator $\hat{J}^2=\hat{J}_x^2+\hat{J}_y^2+\hat{J}_z^2$, with eigenvalue $J(J+1)$, and of the magnetization operator $\hat{J}_z$, with eigenvalue $M$. Here the collective spin operator $\hat{J}_\alpha \equiv \frac{1}{2}\sum_{n=1}^{N}\hat{\sigma}_{\alpha,n}$ is defined as the sum of all $N$ spin operators along the direction $\alpha=\{x,y,z\}$. In an unresolvable spin ensemble, $J$ and $M$ remain good quantum numbers because $\hat{J}^2$ and $\hat{J}_z$ are invariant under particle permutations. The degeneracy index $i=1,2,\dotsc,d^J_N$ labels one of the $d^J_N$ distinct ways to construct a state with the same $J$, where
\begin{equation}
    d^J_N=\frac{N!(2J+1)}{\left(\frac{N}{2}-J\right)!(\frac{N}{2}+J+1)!}, \label{degeneracy: main text}
\end{equation}
gives the total number of degeneracies. We note that two states with different values of $i$ are orthonormal even if they have the same $J$ and $M$, i.e., $\braket{J,M,N,i}{J,M,N,i'} = \delta_{i,i'}$. Appendix \ref{Appendix:rise} provides a detailed discussion on how the degeneracy index $i$ arises from the rules of angular momentum addition.

Mathematically, the lack of resolvability implies that we have no knowledge about the degeneracy index $i$, so a general PI state is an equal mixture of all $d^J_N$ degenerate states, given by \cite{PI_definition1, PI_definition2, Nori}
\begin{equation}
\begin{split}
    \rho_{\text{PI}} =\sum_{J=J_{\text{min}}}^{N/2} p_J \frac{1}{d^J_N} \sum_{M,M'=-J}^{J}\sum_{i=1}^{d^J_N}\rho_{M,M'}^J\\
    \times \ket{J,M,N,i}\bra{J,M',N,i},
    \label{General PI state: introduction}
\end{split}
\end{equation}
where $J_{\text{min}}=0$ for even $N$ and $1/2$ for odd $N$, $p_J$ is the probability of finding the ensemble in different $J$ subspaces, and $\rho^J_{M,M'}$ are matrix elements which are independent of $i$. We note that PI states do not allow coherences between states with different $J$.

In this work, we focus on logically pure PI states where all collective properties are known, i.e., the total angular momentum $J$ and the coherences between different magnetization $M$ are known. These states arise in many situations of practical interest. For example, they can be prepared deterministically through interactions with an ancilla that couples uniformly to all spins \cite{PI_preparation_deterministic1, PI_preparation_deterministic2}, or be stabilized by engineered dissipation \cite{Ruchir}. They can also be prepared probabilistically when ensemble's information has been excessively learned through a collective measurement \cite{PI_preparation_probabilistic1, PI_preparation_probabilistic2, PI_preparation_specific1, PI_preparation_specific2}. One may start with a mixture of different $J$ values, apply phase estimation using an additional ancilla to learn $J$ \cite{Terhal_Dicke_preparation}, and then pump the system to the maximum or minimum $M$ to generate a logically pure PI state. Although these states are physically mixed, as long as their mixedness arises solely from the practical unresolvability of individual spins, not from decoherence or noise, they fall within our framework. Mathematically, this implies two conditions: (1) the state must reside in a single $J$ subspace, i.e., $p_J=\delta_{J,J'}$ for some known $J'$; and (2) $\rho_{M,M'}^J$ is a rank-1 matrix, meaning that the state remains pure in the irreducible representation \cite{Harsh}.

Under these conditions, logically pure PI states take the form
\begin{equation}
    \rho_{\text{PI}}^J=\sum^{d^J_N}_{i=1}\frac{1}{d^J_N}\ketbra{\psi_{i}^J},
    \label{Superposition PI state}
\end{equation}
where
\begin{equation}
    \ket{\psi_{i}^J}=\sum_{M=-J}^{J} c_M\ket{J,M,N,i}.
    \label{Superposition PI state: components}
\end{equation}
The superposition amplitudes $c_M$ are defined by $\rho^J_{M,M'}\equiv c_M c_{M'}^*$, and they are independent of $i$. A pictorial representation of such states is shown in Fig.~\ref{Setup}b. As will become clear in Sec.~\ref{Subsection: Proof of exact quantification}, the first condition ensures that the degeneracy index $i$ can always be determined by local measurements. This enables us to exactly calculate the entanglement of formation by averaging the entanglement of each $i$ state. The second condition ensures that each $i$ state is pure, so that their entanglement can be quantified by the standard entanglement entropy.

\subsection{\label{Subsection: LOCC distinguishability of degenerate basis states}LOCC distinguishability of degenerate basis states}
Our exact entanglement quantification relies on the crucial property that the degeneracy index $i$ can be identified by local operations and classical communication (LOCC) while preserving the superposition in Eq.~(\ref{Superposition PI state: components}). To see this, suppose the $N$-spin ensemble is split into two subensembles 1 and 2, each containing $n$ and $N-n$ spins respectively. Without loss of generality, we assume $n \leq N/2$, so subensemble 1 always has fewer spins. Following the standard angular momentum addition rules \cite{Sakurai}, any degenerate basis state $\ket{J,M,N,i}$ can be written as a superposition of the tensor product of the degenerate basis states for the two subensembles $\left\{\ket{j_1,m_1,n,i^{j_1}}\otimes \ket{j_2,m_2,N-n,i^{j_2}}\right\}$:
\begin{equation}
\begin{split}
    \ket{J,M,N,i}=&\sum_{m_1=-j_1}^{j_1}\sum_{m_2=-j_2}^{j_2} \leftindex^{J,M}C_{j_1,m_1;j_2,m_2}\\
    &\times\ket{j_1,m_1,n,i^{j_1}}\otimes \ket{j_2,m_2,N-n,i^{j_2}},
    \label{Basis state being split with i}
\end{split}
\end{equation}
where $j_k$, $m_k$, and $i^{j_k}$, respectively, represent the total angular momentum, magnetization, and degeneracy index of the subensemble $k=1,2$; $\leftindex^{J,M}C_{j_1,m_1;j_2,m_2}=\bra{j_1,m_1;j_2,m_2}\ket{J,M}$ are the Clebsch-Gordan (CG) coefficients. For non-zero CG coefficients, the subensembles' parameters $\{j_1,\,j_2,\,m_1,\,m_2\}$ obey the rules of addition of angular momenta: (I) $|j_1-j_2|\leq J\leq (j_1+j_2)$, and (II) $m_1+m_2=M$.

Since many pairs of $(j_1,j_2)$ can satisfy rule (I) for a given $J$, and each pair has $d^{j_1}_{n}d^{j_2}_{N-n}$ degeneracies, knowing only $j_1$ and $j_2$ is insufficient to distinguish between different $i$ states. Fortunately, states with the same $J$ but different $i$ must be composed of subensemble states with a different combination of subensemble degeneracy indices \cite{PI_definition1}, $i^{j_1}$ and $i^{j_2}$. If one can locally identify $i^{j_1}$ and $i^{j_2}$, the degeneracy index $i$ is revealed. In fact, such a set of local measurements exist. Since $\left\{\ket{j_1,m_1,n,i^{j_1}}\right\}$ and $\left\{\ket{j_2,m_2,N-n,i^{j_2}}\right\}$ are respectively orthonormal bases for subensembles 1 and 2, $i^{j_1}$ and $i^{j_2}$ can be identified by locally measuring the subensembles with the measurement operators that are identities in each subensemble degenerate subspace, i.e., $\left\{\hat{M}_{i^{j_k}}=\sum_{m_k}\ketbra{j_k,m_k,N_k,i^{j_k}}\right\}$. Since PI states do not involve any superposition between states with different degeneracy indices, these measurements will only reveal the degeneracy index $i$ while preserving the superposition between different $M$ eigenstates, i.e., $\hat{M}_{i^{j_1}}\otimes \hat{M}_{i^{j_2}} \ket{\psi^J_i}=\ket{\psi^J_i}$. This implies that the degenerate basis states are distinguishable via LOCC.

We note that our formalism does not require these local measurements to be performed experimentally; it suffices that they exist in principle. As will become clear in Sec.~\ref{Subsection: Proof of exact quantification}, their existence allows us to place a meaningful bound on the entanglement of formation.

\section{\label{Section: Exact quantification of entanglement for PI states}Exact quantification of entanglement for PI states}
\subsection{\label{Subsection: Entanglement of formation and entropy of entanglement}Entanglement of formation and entanglement entropy}
Among different entanglement measures, here we choose entanglement of formation $E_F$ due to its relevant operational meaning. Explicitly, entanglement of formation quantifies the entanglement cost of preparing $\rho$: it equals the asymptotic number of maximally entangled qubits required to create the state under LOCC \cite{Entanglement_measures}.

Entanglement of formation for a mixed state $\rho$ is defined as the infimum average entanglement over all possible decomposition of the state \cite{Entanglement_measures}:
\begin{equation}
    E_F(\rho)\equiv \underset{\{p_k,\ket{\psi_k}\}}{\inf} \sum_k p_k E(\ket{\psi_k}),
    \label{entanglement of formation}
\end{equation}
where $\rho=\sum_k p_k \ketbra{\psi_k}$ represents the ensemble state as a statistical mixture of many pure states $\ket{\psi_k}$ with probability $p_k$. $E$ is the entanglement entropy, defined by \cite{Characterizing_entanglement_for_symmetric_systems, Entanglement_measures}
\begin{equation}
   E(\ket{\psi_k})=-\sum_{\alpha} \lambda_{\alpha}\log_2{\lambda_{\alpha}},
    \label{Von Neumann entropy of entanglement}
\end{equation}
where $\lambda_{\alpha}$ are the eigenvalues of the reduced density matrix $\rho_1=\Tr_{2}{\ketbra{\psi_k}}$.

In general, entanglement of formation is extremely difficult to calculate because there can be infinitely many possible decompositions, and finding the infimum is challenging, if at all possible. So far, we only know the analytical expression of the entanglement of formation in very limited cases, such as two-qubit mixed states \cite{Wooter}. We will show that the decomposition in Eq.~(\ref{Superposition PI state}) is the one that minimizes the average entanglement in Eq.~(\ref{entanglement of formation}) for all logically pure PI states, so their entanglement of formation is exactly the average entanglement entropy of the degenerate states. Our result adds a huge class of PI states to the list of states whose entanglement of formation is exactly computable.

\subsection{\label{Subsection: Proof of exact quantification}Proof of exact quantification}
We start by rephrasing Eq.~(\ref{entanglement of formation}). Generally, for any choice of $\{p_k, \ket{\psi_k}\}$ that yields a mixed state $\rho$, the average entanglement entropy of the components should be no smaller than the entanglement of formation \cite{Entanglement_measures},
\begin{equation}
    E_F(\rho)\le \sum_k p_k E(\ket{\psi_k}).\label{Lower bound}
\end{equation}

Now, we exploit the fact that entanglement is a non-local quantum resource that cannot increase under local operations. Consider a decomposition where the components $\{\ket{\psi_k}\}$ can be distinguished and not destroyed by locally measuring the subensembles. After measurement, the outcome $\rho_k=\ketbra{\psi_k}$ is obtained with probability $p_k$. Since entanglement of formation is an LOCC monotone \cite{Entanglement_measures}, the average entanglement of the post-measured state is upper bounded by the entanglement of formation
\begin{equation}
    \sum_k p_k E(\ket{\psi_k}) \le E_F(\rho). \label{Upper bound}
\end{equation}

Comparing Eqs.~(\ref{Lower bound}) and (\ref{Upper bound}), we see that the entanglement of formation for the mixed states whose components can be distinguished by local non-demolition measurements is given exactly by
\begin{equation}
    E_F(\rho)=\sum_k p_k E(\ket{\psi_k}).\label{Exact entanglement}
\end{equation}
Equation (\ref{Exact entanglement}) is the basis of our scheme. As discussed in Sec.~\ref{Subsection: LOCC distinguishability of degenerate basis states}, states with different degeneracy indices can be distinguished and preserved by measuring the degeneracy indices of each subensemble. As a result, for a logically pure PI state, we have
\begin{equation}
    E_F(\rho_{\text{PI}}^J)=\sum_i p_i E(\ket{\psi^J_i}).
    \label{Exact entanglement for PI state}
\end{equation}

Before moving forward, we discuss several points related to the measurement mentioned above. First, although measurement usually disturbs states, here our measurement of subensemble degeneracy indices will preserve the PI state. This can be understood by observing that the initial state can be restored simply by forgetting the measurement outcomes, i.e., the degeneracy index. Second, the degenerate basis states $\ket{J,M,N,i}$ are generally not PI, nor the states $\ket{j_1,m_1,n,i^{j_1}}$ and $\ket{j_2,m_2,N-n,i^{j_2}}$ in each subensemble. The consideration of non-PI measurements is however not contradictory to the studies of unresolvable spin ensembles. The resolution is that the measurement operators are only theoretical tools introduced to prove that the decomposition in Eq.~(\ref{Superposition PI state}) minimizes the average entanglement entropy. Our method quantifies the entanglement in the subensembles even if the spins are resolvable. For unresolvable spins, it gives an upper bound to the entanglement that can be extracted. \cite{Superselection_rule}.

Lastly, although our method is exact only for the PI states in Eq.~(\ref{Superposition PI state}) whose collective properties are known, it can also be used as a non-trivial entanglement upper bound for the most general PI states in Eq.~(\ref{General PI state: introduction}):
\begin{equation}
    E_F\left(\rho_{\text{PI}}\right) \le \sum_{J,k} p_J \mu_k^J \frac{1}{d^J_N} \sum_i E\left(\ket{\psi_{k,i}}\right),
    \label{upperbound}
\end{equation}
where $\mu_k^J$ and $\ket{\psi_{k,i}}$ are the $k$th eigenvalue and eigenvector of $\rho^J_{M,M'}$ which is generally not rank-1. In this work, we focus on the exact quantification. Experimental validation of Eq.~(\ref{upperbound}) is generally nontrivial and will be studied in future work.

\subsection{\label{Subsection: Formula for entanglement of PI states}Formula for entanglement of PI states}
Based on Eq.~(\ref{Exact entanglement}), the entanglement of formation of a PI state in the form of Eq.~(\ref{Superposition PI state}) with partition number $n$ is exactly given by
\begin{align}
    E_F(\rho^J_{\text{PI}};n)=&-\frac{1}{d^J_N}\sum_{j_1=j_{1,\text{min}}}^{n/2}\sum_{\substack{j_2=j_{2,\text{min}}\\}}^{(N-n)/2} 
    d^{j_1}_n d^{j_2}_{N-n} \nonumber \\
    &\times\sum_k \nu_{k}^{(j_1,j_2)} \log_2{\nu_{k}^{(j_1,j_2)}}.
    \label{Gold equation for entanglement: Superposition}
\end{align}
The derivation is detailed in Appendix \ref{Appendix:derivation}. The first two summations account for all possible combinations of $j_1$ and $j_2$ that satisfy Rule II; for those that do not satisfy Rule II, their contribution is simply zero due to the vanishing CG coefficients. For each combination, there are respectively $d^{j_1}_n$ and $d^{j_2}_{N-n}$ degenerate states in subensembles 1 and 2 that have the same amount of entanglement. The last summation computes the entanglement entropy of each degenerate state according to Eq.~(\ref{Von Neumann entropy of entanglement}), where $\nu_{k}^{(j_1,j_2)}$ are the eigenvalues of the reduced density matrix of subensemble 1:
\begin{widetext}
    \begin{equation}
\begin{split}
    \sigma_1^{(j_1,j_2)}=&\sum_{M=-J}^{J} \sum_{M'=-J}^{J}\rho^J_{M,M'}\sum_{m_1=-j_1}^{j_1} \sum_{m_1'=-j_1}^{j_1}
    \prescript{J,M}{}{C_{j_1,m_1;j_2,M-m_1}}
    \prescript{J,M'}{}{C_{j_1,m_1';j_2,M-m_1}} \ketbra{j_1, m_1}{j_1, m_1'}.   \label{reduced density matrix for superposition}
\end{split}
\end{equation}
\end{widetext}
This expression is obtained by taking the partial trace of $\ketbra{\psi_i^J}$ over subensemble 2. Here, we omit the subensemble degeneracy index because all degenerate states with the same $(j_1,j_2)$ share the same entanglement entropy.

\section{\label{Section: Angular momentum eigenstates}Angular momentum eigenstates}
We first look at the simultaneous eigenstates of total angular momentum and magnetization, i.e.,
\begin{equation}
    \rho_{\text{PI}}^{J,M}=\sum^{d^J_N}_{i=1}\frac{1}{d^J_N}\ketbra{J,M,N,i}.
    \label{No superposition PI state}
\end{equation} 
Because every component $\ket{J,M,N,i}$ is already in the Schmidt form (Eq.~(\ref{Basis state being split with i})), the Schmidt coefficients are simply the CG coefficients squared. Thus, the entanglement of formation is reduced to
\begin{widetext}
    \begin{equation}
        E(\rho_{\text{PI}}^{J,M};n)=-\frac{1}{d^J_N} \sum_{j_1=j_{1,\text{min}}}^{n/2}\sum_{\substack{j_2=j_{2,\text{min}}\\}}^{(N-n)/2} d^{j_1}_{n}d^{j_2}_{N-n} \sum_{m_1=-j_1}^{j_1}\left(^{J,M}C_{j_1,m_1;j_2,M-m_1}\right)^2 \log_2{\left(^{J,M}C_{j_1,m_1;j_2,M-m_1}\right)^2}. \label{Gold equation for entanglement}
    \end{equation}
\end{widetext}

In the following, we will use this formula to survey the change of entanglement against different collective ensemble parameters. Particularly, we will compare the behaviors of PI states with different $J$ to that of Dicke states ($J=N/2$), which is the only class of PI state that is pure and well-studied.

\subsection{Entanglement vs. magnetization $M$}

\begin{figure*}[t!]
    \centering
    \includegraphics[width=17.9cm]{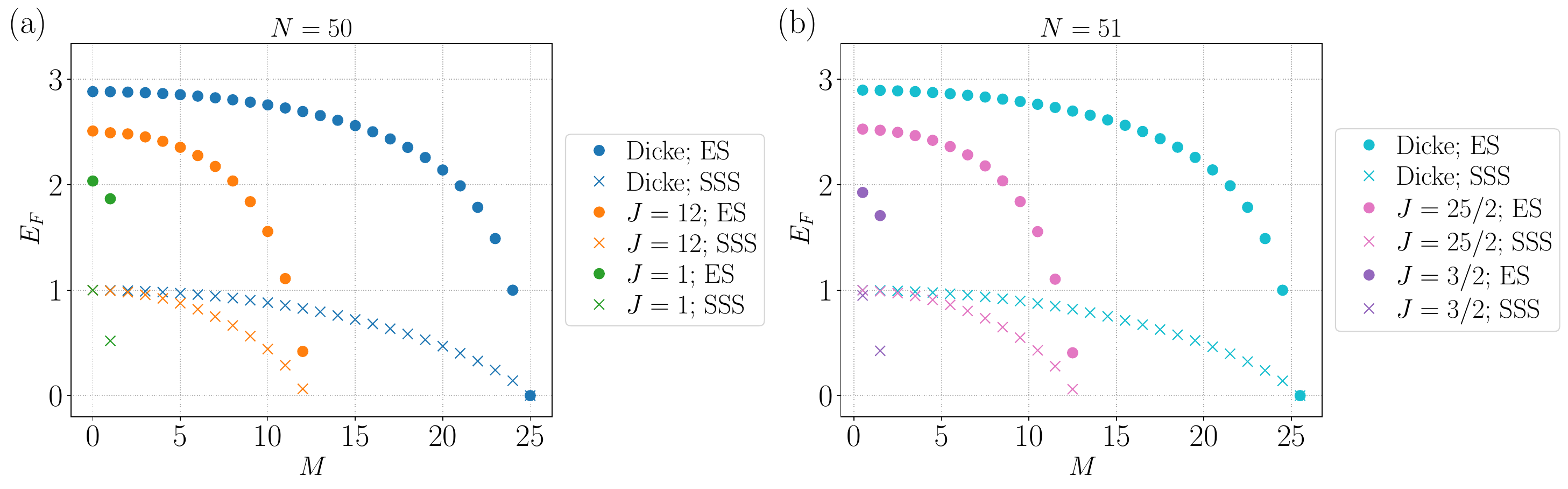}
    \caption{$E_F$ vs. $M$ for (a) $N=50$ (even) and (b) $N=51$ (odd), shown in both splitting scenarios. We only show the data for non-negative $M$, because the states with $-M$ have exactly the same amount of entanglement as those with $+M$.}
    \label{EM plot}
\end{figure*}

We start by exploring how entanglement depends on magnetization $M$. We consider two splitting scenarios: single-spin splitting (SSS), where one subensemble contains only $n=1$ spin, and even splitting (ES), where the spins are most evenly distributed among the subensembles, i.e., $n=\lfloor N/2\rfloor$. We note that $\rho_{\text{PI}}^{J,M}$ and $\rho_{\text{PI}}^{J,-M}$ have the same entanglement because they are interconvertible by individually flipping all the spins, so we focus on states with $M \geq 0$.

In Fig.~\ref{EM plot}, we show the typical results for an ensemble. We computed the entanglement of magnetization eigenstates for different total angular momentum $J$. Across all tested cases, states with different $J$ generally exhibit behavior similar to Dicke states, whose entanglement is monotonically decreasing as magnetization $|M|$ increases. This trend arises because states with larger $|M|$ can be composed of fewer combinations of $m_1$ and $m_2$ according to Rule II, and so the number of superpositions in Eq.~(\ref{Basis state being split with i}) decreases. Generally, having more superposition components means more entanglement because the more non-zero CG coefficients in Eq.~(\ref{Basis state being split with i}), the higher the entanglement according to Eq.~(\ref{Gold equation for entanglement}). Due to this general monotonic behavior, our subsequent analysis will mainly study maximum-$|M|$ ($M=J$) and minimum-$|M|$ ($M=0,1/2$) states as the representative cases.

However, we observe a key difference between Dicke states and other states. For Dicke states, the maximum magnetization states ($M = J$) are separable, whereas for general $J$, they must be entangled. The main reason is that when $J\neq N/2$, some degenerate components can have $j_1 + j_2 > J$. In such cases, the $M = J$ state involves a superposition of multiple states with different $m_1$ and $m_2$, so there will be entanglement. This is not the case for $J=N/2$, where $j_1+j_2 = J$ and there is only one component with $m_1 = j_1$ and $m_2=j_2$.

Moreover, for single-spin splitting, all the states with $M=0$ are maximally entangled while those with $M=1/2$ are not. This is due to the fact that $M=0$ requires the states to be symmetric under a global spin flip, so the CG coefficients must be equal, leading to maximal entanglement. We will explain this more thoroughly in the next subsection.

\subsection{\label{Entanglement vs. total angular momentum}Entanglement vs. total angular momentum $J$}
As shown in the previous subsection that some properties of entanglement depend on total angular momentum $J$, here we want to look at such dependence in greater detail. Due to the general monotonic behavior in magnetization, we consider two representative cases: the maximum-$M$ states ($M=J$) and the minimum-$|M|$ states ($M=0,1/2$).

\begin{figure*}[t!]
    \centering
    \includegraphics[width=17cm]{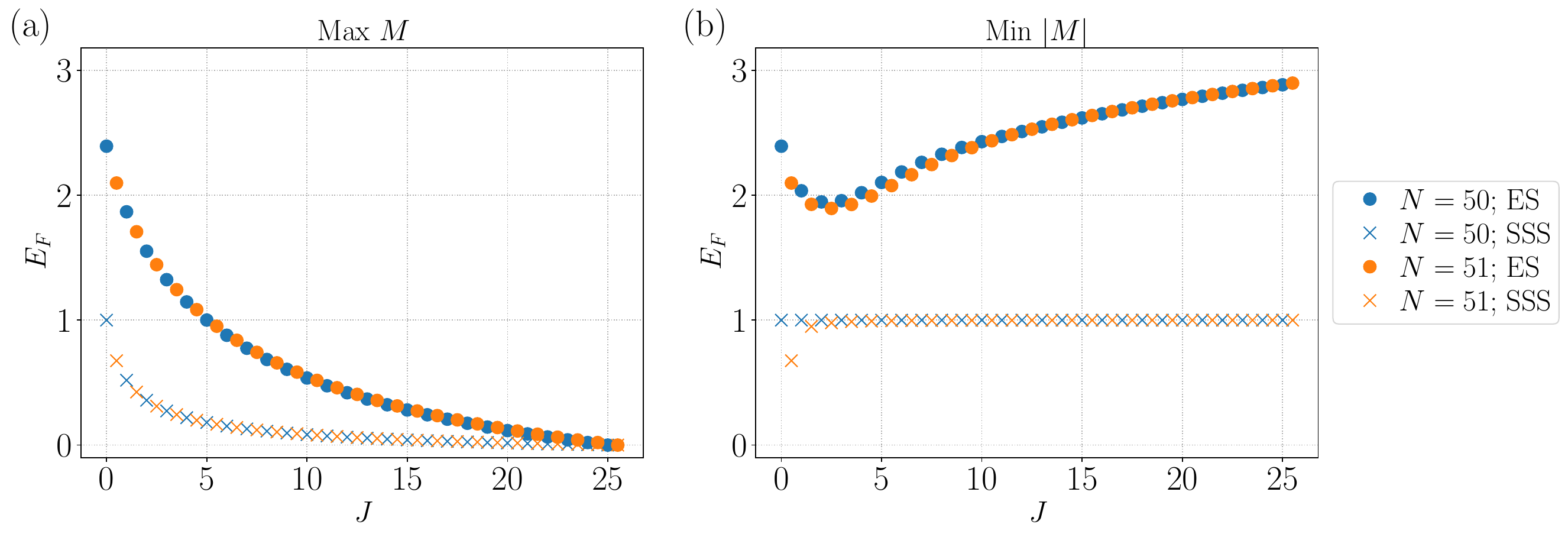}
    \caption{$E_F$ vs. $J$ for (a) $M=J$ and (b) $M=0,1/2$ states, shown in both splitting scenarios.}
    \label{EJ plot M=J}
\end{figure*}

In Fig.~\ref{EJ plot M=J}a, we show a typical result for maximum-$M$ states. The entanglement is monotonically decreasing for both even splitting and single-spin splitting. The parity of $N$ does not seem to play a role here. The monotonic behavior can be explained by the same reasoning as in the previous section that there are more combinations of $j_1+j_2>J$ for lower $J$. For single-spin splitting, because one subensemble contains only one spin, we can obtain a simple expression for the entanglement (see Appendix \ref{Appendix:single})
\begin{equation}
\begin{split}
    &E_F(\rho_{\text{PI}}^{J,J};n=1)\\
    =&\frac{1}{2}\left(1-\frac{J}{\frac{N}{2}}\right)\frac{1}{2J+1}\\
    &\times\left\{(2J+2)\log_2{(2J+2)}-(2J+1)\log_2{(2J+1)}\right\},
    \label{Entanglement n=1 M=J}
\end{split}
\end{equation}
which is indeed a monotonic function of $J$ that vanishes at $J=N/2$.

In contrast, the entanglement behavior of the minimum-$|M|$ states in Fig.~\ref{EJ plot M=J}b is more complicated. Firstly, the patterns in the two splitting scenarios are strikingly different. For even splitting, there is a dip of entanglement. Intuitively, it can be understood from the fact that the two subensembles are strongly anti-correlated at minimum total angular momentum $J=0,1/2$ and correlated at maximum $J=N/2$. Since entanglement is a measure of quantum correlation, we expect it will be the highest in these two scenarios, where the (anti-)correlation is the strongest. Conversely, the correlation will be weak at some intermediate $J$, so a dip in entanglement is expected. However, we evaluated the correlation of the ensemble state, $|\langle \vec{j}_1 \cdot \vec{j}_2 \rangle|$, and found that the value of $J$ at which the correlation is minimized does not align with the entanglement dip. This suggests that the entanglement also depends on other properties of the subensemble states apart from correlation.

\begin{figure}[t!]
    \centering
    \includegraphics[width=8cm]{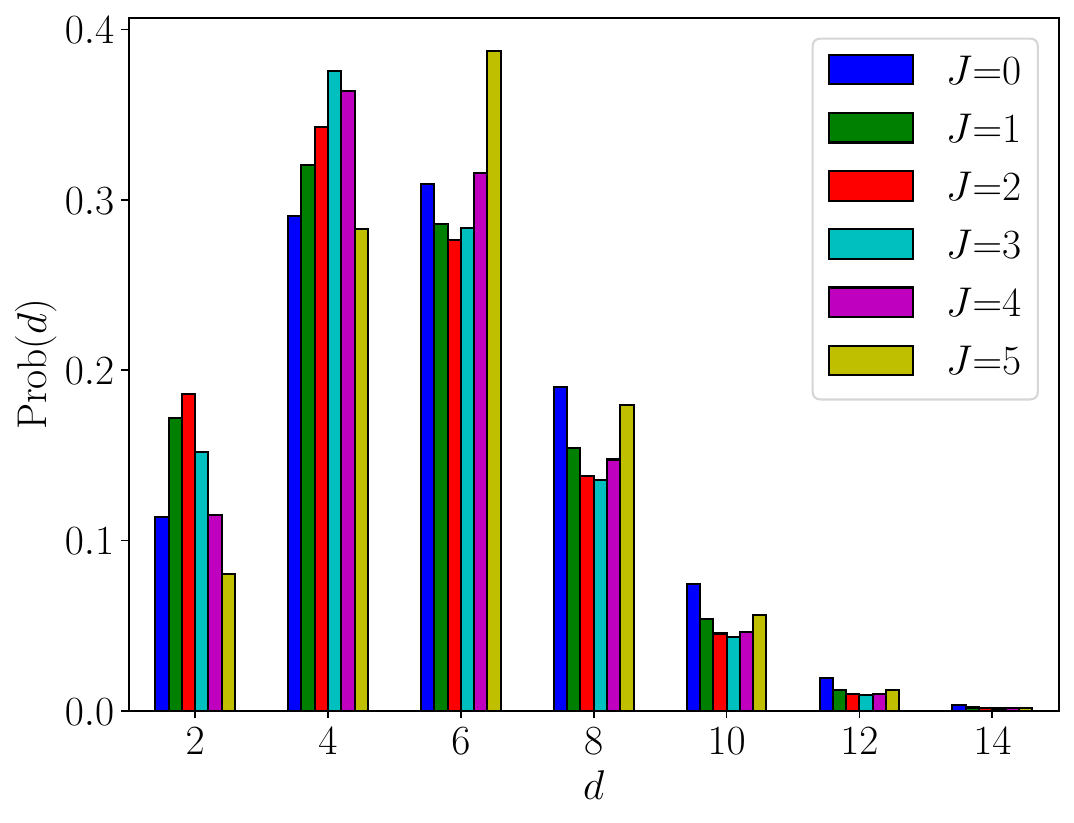}
    \caption{Probability distribution components with qu$d$it levels $d=2\min(j_1,j_2)+1$, $\text{Prob}(d)=\sum_{\{j_1,j_2|d=2\min{(j_1,j_2)+1}\}}d^{j_1}_n d^{j_2}_{N-n}/d^J_N$ for the six lowest $J$ states with $N=50$ and $M=0$. The distribution is plotted up to $d=14$ since higher values have negligible probabilities. Components with larger $d$ permit more levels for superposition and hence entanglement. The population of low $d$ states $(d=2,4)$ peaks when $J=2,3$.}
    \label{d_distribution}
\end{figure}

To look into this matter further, we first recall from Eq.~(\ref{Basis state being split with i}) and Rule II that a component with subensemble angular momenta $j_1$, $j_2$ is a pair of entangled qu$d$its, where $d = 2\min(j_1,j_2)+1$. Although the precise amount of entanglement of each component depends on the superposition amplitudes, which are related to the CG coefficients, a larger $d$ will generally allow for more entanglement. Therefore, we conjecture that the average entanglement is low when the PI state is composed of a high portion of qu$d$it pairs with low $d$. To verify this, we plot the probability distribution of $d$ for a sample case with $N=50$ in Fig.~\ref{d_distribution}. We see that the probability of having a component with low $d$ (i.e., $d=2,4$) peaks for states with $J=2$ and $3$. This agrees with our observation in Fig.~\ref{EJ plot M=J}b that entanglement is minimum around $J=2$ and $3$.

On the other hand, for single-spin splitting, no dip of entanglement is observed. More intriguingly, the entanglement behavior shows a strong dependence on the parity of the total spin number $N$: for even-$N$, the two subensembles are always maximally entangled, i.e., $E_F =1$; for odd $N$, entanglement rises as $J$ increases and approaches to maximal entanglement. To understand the maximal entanglement in even $N$, we recall that subensemble 1 is a single spin and subensemble 2 contains odd number of spins. Since the minimal $M$ for even $N$ is $M=0$, every component state should be symmetric against flipping all the spins. As a result, each component state is an equal superposition of $\ket{1/2,-1/2}\ket{j_2,1/2}$ and $\ket{1/2,1/2}\ket{j_2,-1/2}$, which is equivalent to a maximally entangled qubit pair. However, there is no such symmetry for odd-$N$ minimal-$|M|$ states, which have $M=1/2$. Therefore the superposition amplitudes of $\ket{1/2,1/2}\ket{j_2,0}$ and $\ket{1/2,-1/2}\ket{j_2,1}$ are not necessarily equal. The monotonic rising trend, though, originates from the subtle behavior of the CG coefficients, as presented in Appendix \ref{Appendix:single}.

\subsection{Entanglement vs. particle number $N$}
We have seen that the entanglement properties can depend on the number of spins $N$, in particular on its parity. Here, we investigate further into such dependence.

\begin{figure*}[t!]
    \centering
    \includegraphics[width=17.4cm]{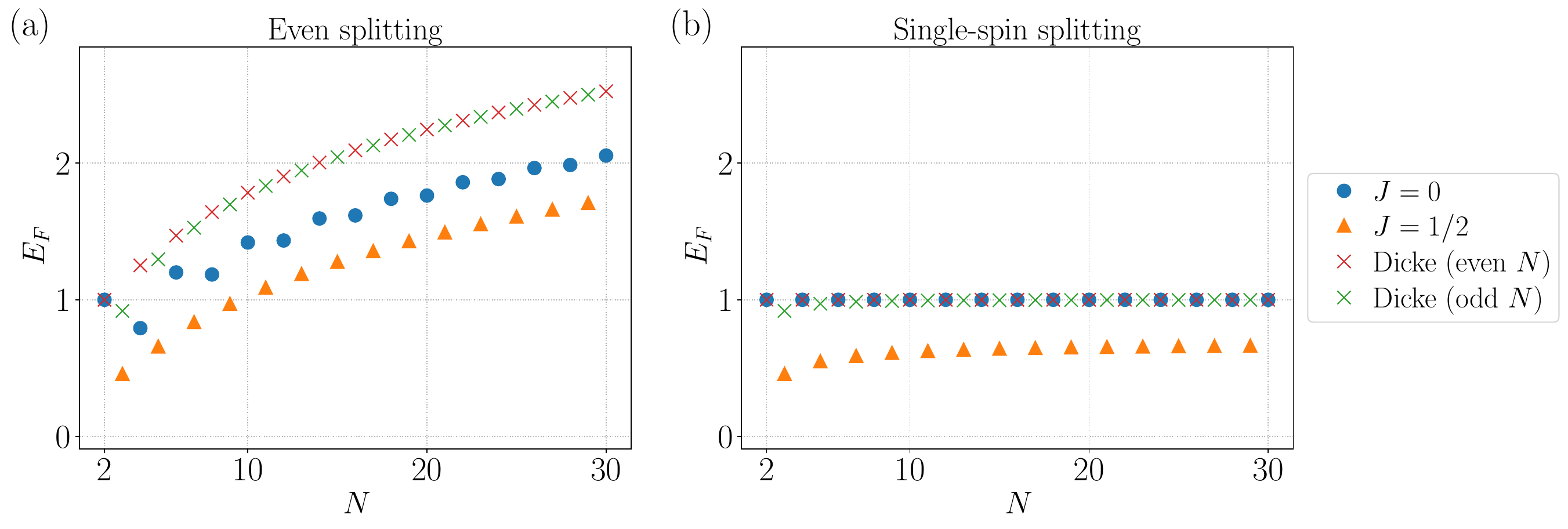}
    \caption{$E_F$ vs. total spin number $N$ with minimum $|M|$ for (a) even splitting and (b) single-spin splitting. We compare the lowest $J$ states ($J=0$ for even $N$ and 1/2 for odd $N$) with Dicke states ($J=N/2$).}
    \label{EN min M plot}
\end{figure*}

In Fig.~\ref{EN min M plot}, we compare the entanglement of the two lowest $J$ states with that of the Dicke states. One can see that entanglement generally increases with $N$. This is because a larger $N$ will introduce components with higher $j_1 \leq N/2$ and $j_2 \leq N/2$ in the subensembles. Since each component can be a superposition of subensemble magnetization states $-j_1\leq m_1 \leq j_1$, $-j_2 \leq m_2 \leq j_2$, larger $j_1$ and $j_2$ allow for more superposition and thus higher entanglement.

Surprisingly, for even splitting, the $J=0$ state exhibits a zigzag pattern: entanglement decreases or increases gently as $N$ changes from odd to even multiples of 2, but increases sharply when going from even to odd multiples of 2. This behavior suggests that entanglement depends on the parity of the subensemble spin number $n = N/2$. We will discuss the cause of this phenomenon in the next section.

One may also be interested in the scaling of entanglement in the classical limit, i.e., as $N\rightarrow\infty$. For even splitting, the maximum dimension of each subsystem is $j_1=j_2=N/4$, which is the case for Dicke state, so the maximum entanglement is equivalent to that of a maximally entangled qu$d$it with $d=N/2+1$. This suggests that the maximum entanglement scales as $\log{N}$, so entanglement per spin $E_F/N$ vanishes in this classical limit.

For single-spin splitting, the even-$N$ states are always maximally entangled because the bipartition always results in $j_1=1/2$ and $j_2$ being a half-integer. Because $M=0$, the states are invariant against the flipping of all spins. The superposition amplitudes of the states containing $m_1 = \pm1/2$ are thus equal, which results in a maximally entangled state. In contrast, the entanglement of odd-$N$ states varies as $N$. It is because these PI states can be composed of two types of components, either $j_2 = J+1/2$ or $j_2 = J-1/2$. According to Eq.~(\ref{degeneracy: main text}), the ratio of these components varies as $N$. Consequently, the entanglement in each component state also depends on $N$ in general.

\subsection{Entanglement vs. partition number $n$}
\begin{figure*}[t!]
    \centering
    \includegraphics[width=17.9cm]{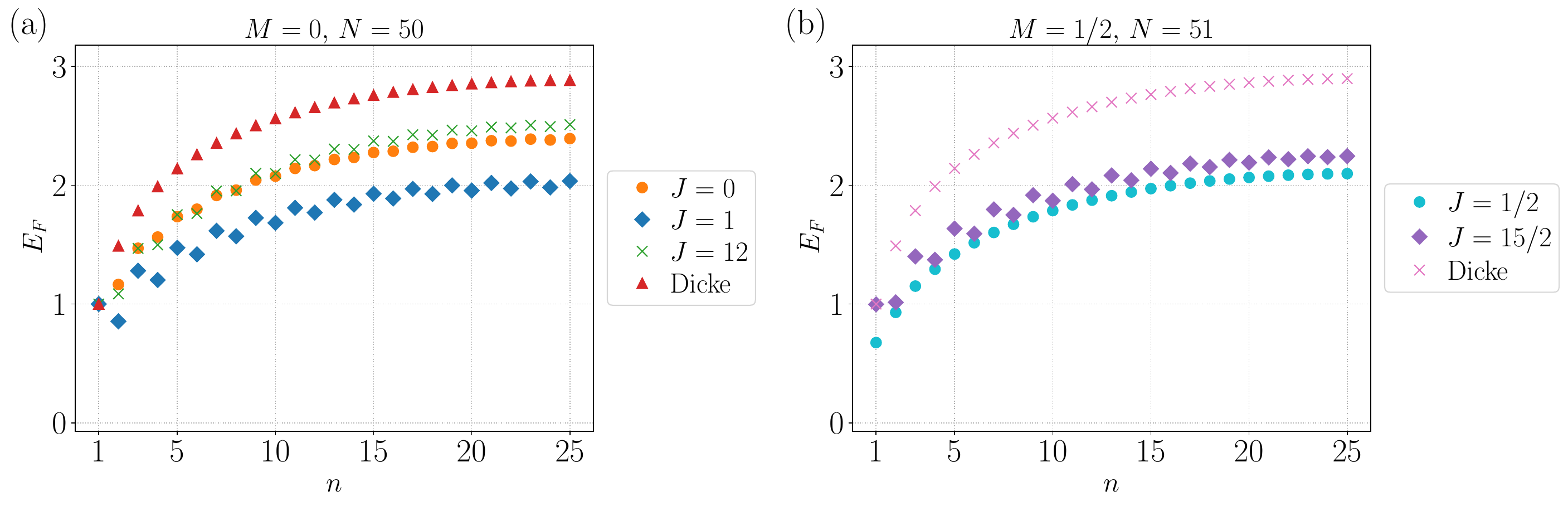}
    \caption{$E_F$ vs. partition number $n$ for minimum-$|M|$ states, shown for (a) $N=50$ with integer $J$ and (b) $N=51$ with half-integer $J$. Some low-$J$ states exhibit a zigzag pattern as $n$ varies.}
    \label{En plot}
\end{figure*}

Inspired by the zigzag pattern observed in the last section, here we study the entanglement dependence on the partition number $n$. In Fig.~\ref{En plot}, we present the typical behavior for an ensemble with even $N$ and odd $N$. As we can see, entanglement generally rises as $n$ increases. It is because a larger $n$ generally allows components with higher $j_1 \leq n/2$, so the states can involve more superposition of $m_1$ states, and thus contain more entanglement.

Intriguingly, we observe zigzag pattern for states with low $J$ and $M$. This is in stark contrast to large $J$ states, including the Dicke states, where entanglement increases monotonically as $n$. The origin of the zigzag pattern is slightly different for the ensembles with even $N$ and odd $N$, so we will discuss them separately.

For even $N$, $j_1$ and $j_2$ are both half-integers when $n$ is odd, while they have to be integers when $n$ is even. In the former case, an integer-$M$ state must be composed of a superposition of half-integers $m_1$ and $m_2$ states, so every component is entangled. On the contrary, since $j_1$ and $j_2$ are integers when $n$ is even, an integer-$M$ state can be composed of components that one of the subensembles is in the singlet state, i.e., either $j_1=0$ or $j_2=0$. Such components contain no superposition and are separable---thereby reducing the average entanglement. This also explains why the zigzag pattern is missing in Dicke states, because they consist of only one component with $j_1=n/2$ and $j_2=(N-n)/2$, so no singlet can be formed.

Surprisingly, the zigzag pattern also appears for odd $N$, in which one subensemble must contain an even number of spins while the other is odd. To explain this, we first consider when $n$ is odd. In this case, $j_1$ is a half-integer, so $j_2$ must be a non-zero integer when $J>1/2$. Under these conditions, no singlet can form, and all components are entangled. Conversely, when $n$ is even, $j_1$ can be zero as long as $J\leq(N-n)/2$. These components are separable because subensemble 1 will form a singlet state. As a result, the average entanglement for odd $n$ can be higher than that for even $n$. We note that our theory predicts that the zigzag pattern does not appear when $J$ is either too large or at the minimum $J=1/2$. This agrees with the observation in Fig.~\ref{En plot}b.

\section{\label{Section: Superposition states}Superposition states}
In this section, we extend our studies to the PI states that are a superposition of magnetization eigenstates (see Eq.~(\ref{Superposition PI state})). Entanglement for these states can be calculated by Eq.~(\ref{Gold equation for entanglement: Superposition}), which is equal to the average entanglement of every degenerate component $\ket{\psi^J_i}$. We will apply our method to specifically study two classes of superposition states that could offer quantum advantages in metrology: GHZ states and spin-squeezed states.

Before we show the explicit results, we want to discuss how the magnetization superposition can lead to richer entanglement behavior. We can see that the entanglement of these states comes from two main contributions: the ``intrinsic'' superposition in each magnetization eigenstate and the ``extrinsic'' superposition of different magnetization eigenstates. Specifically, consider a particular degenerate component,
\begin{equation}
    \ket{\psi_i^J}=\sum_M c_M \sum_{m_1,m_2} \leftindex^{J,M}C_{j_1,m_1;j_2,m_2} \ket{j_1,m_1}\ket{j_2,m_2}.
    \label{Superposition state discussion}
\end{equation}
If each subensemble component $\ket{j_1,m_1}$ and $\ket{j_2,m_2}$ is associated with only one $M$ state, it is easy to see that the entanglement is a direct sum of the entanglement of each $M$ state and the entanglement entropy, as a result of the superposition amplitudes $c_M$ (see Appendix \ref{Appendix:superposition}). However, this is not always the case, because magnetization eigenstates with different $M$ can be composed of the same component from either of the subensembles. In this situation, the extrinsic and intrinsic superposition will interfere and produce intriguing entanglement behaviors.

\subsection{GHZ-like states}
The GHZ states are genuinely $N$-partite entangled states, defined as the equal superposition of all spins pointing up and all spins pointing down \cite{Metrology1}:
\begin{equation}
    \ket{\text{GHZ}}=\frac{1}{\sqrt{2}}\left(\ket{\frac{N}{2},\frac{N}{2}}+\ket{\frac{N}{2},-\frac{N}{2}}\right).
    \label{GHZ state}
\end{equation}
The GHZ states are of great interest in quantum sensing as they are the $N$-spin states that are most sensitive to collective phase shift.

The GHZ states are PI and reside in the subspace of the maximum total angular momentum, i.e., $J=N/2$. We can consider analogous GHZ states in other $J$ subspaces as equal superposition of the maximum and minimum $M$ states, i.e., $\rho^J_{J,J}=\rho^J_{J,-J}=\rho^J_{-J,J}=\rho^J_{-J,-J}=1/2$ in Eq.~(\ref{General PI state: introduction}), or equivalently $c_J=c_{-J}=1/\sqrt{2}$ in Eq.~(\ref{Superposition PI state: components}), so that
\begin{equation}
    \ket{\text{GHZ},i}=\frac{\ket{J,J,N,i}+\ket{J,-J,N,i}}{\sqrt{2}}.
    \label{GHZ-like state for each i}
\end{equation}

Since the GHZ-like states are a superposition of the maximum and minimum $M$ states, their reduced density matrices are less likely to overlap. Particularly, for $J>N/4$, because $j_1,j_2 \leq N/4$ and $m_1 + m_2 = M$, both $m_1$ and $m_2$ are positive (negative) for $\ket{J,J,N,i}$ ($\ket{J,-J,N,i}$). Therefore, the subensemble states of $\ket{J,J,N,i}$  and $\ket{J,-J,N,i}$ must be different. In these cases, the total entanglement for Eq.~(\ref{GHZ-like state for each i}) can be understood as the sum of the GHZ entanglement and the weighted entanglement of the $M=\pm J$ states. The former always provides one e-bit (entanglement bit), so the behavior of entanglement is dominated by the intrinsic entanglement of the max-$M$ magnetization eigenstates. 

\begin{figure}[htp!]
    \centering
    \includegraphics[width=8cm]{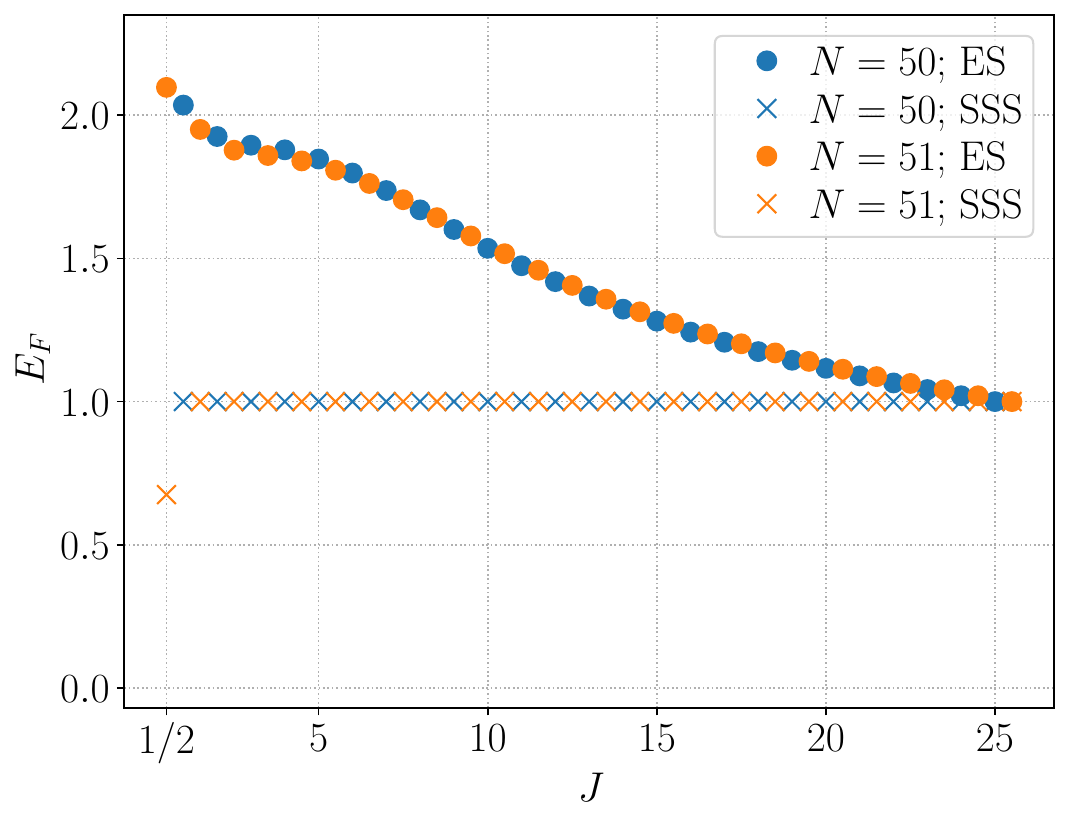}
    \caption{$E_F$ vs. $J$ for GHZ-like states. The value of $J$ ranges from $1/2$ to $N/2$. For single-spin splitting, the GHZ-like states are always maximally entangled, except at $J=1/2$.}
    \label{EJ GHZ states}
\end{figure}

In Fig.~\ref{EJ GHZ states}, we show the entanglement for the GHZ-like states across different $J$. For even splitting, the entanglement decreases with $J$, which largely follows the same trend as the $M=J$ states in Fig.~\ref{EJ plot M=J}a. We note that the $M=-J$ state has exactly the same entanglement as $M=J$ as they are interconvertible by flipping all the spins. The GHZ state entanglement converges to unity at large $J$, instead of vanishing as in Fig.~\ref{EJ plot M=J}a for angular momentum eigenstates. At $J=N/2$ (Dicke subspace), we recover the original GHZ state in Eq.~(\ref{GHZ state}). It possesses exactly one e-bit due to the extrinsic entanglement, while intrinsic entanglement vanishes because $\ket{N/2,N/2}$ and $\ket{N/2,-N/2}$ are respectively all spins pointing up and down, thus separable. For small $J$, the two $M$ states start to overlap, and the resulting behavior becomes more complicated.

For single-spin splitting, the entanglement is always maximum except at $J=1/2$. To understand this behavior, we consider the states' explicit composition. Since $j_1=1/2$ for single splitting, subensemble 2 can only have either $j_2=J+1/2$ or $j_2=J-1/2$. In the former case, every degenerate component can be expressed as
\begin{align}
    \ket{\text{GHZ},i}=&\frac{1}{\sqrt{2}}\ket{\uparrow}\left(\sqrt{\frac{1}{2J+2}}\ket{J+\frac{1}{2},J-\frac{1}{2}}\right. \nonumber \\
    &\qquad+\left.\sqrt{\frac{2J+1}{2J+2}}\ket{J+\frac{1}{2},-J-\frac{1}{2}}\right) \nonumber\\
       -&\frac{1}{\sqrt{2}}\ket{\downarrow}\left(\sqrt{\frac{2J+1}{2J+2}}\ket{J+\frac{1}{2},J+\frac{1}{2}}\right. \nonumber \\
       &\qquad+\left.\sqrt{\frac{1}{2J+2}}\ket{J+\frac{1}{2},-J+\frac{1}{2}}\right).
    \label{GHZ-like states: i=1}
\end{align}
For $J>1/2$, all components inside the two brackets are orthogonal, so the state is already in the Schmidt form. Moreover, because the two Schmidt coefficients are equal to $1/\sqrt{2}$, it is maximally entangled. On the contrary, $J=1/2$ is an exceptional case because the component $\ket{J+1/2,J-1/2}$ is identical to $\ket{J+1/2,-J+1/2}$, so Eq.~(\ref{GHZ-like states: i=1}) is no longer maximally entangled.

Similarly, in the $j_2=J-1/2$ case, the component is given by
\begin{align}
    \ket{\text{GHZ},i}=&\frac{1}{\sqrt{2}}\ket{\uparrow}\ket{J-\frac{1}{2},J-\frac{1}{2}} \nonumber \\
    +&\frac{1}{\sqrt{2}}\ket{\downarrow}\ket{J-\frac{1}{2},-J+\frac{1}{2}}.
    \label{GHZ-like states: i=2}
\end{align}
This state is always maximally entangled whenever $J>1/2$. In the exceptional case $J=1/2$, $\ket{J-1/2,J-1/2}=\ket{J-1/2,-J+1/2}=\ket{0,0}$, so it becomes separable.

\subsection{Spin-squeezed states}

\begin{figure*}[htp!]
    \centering
    \includegraphics[width=17cm]{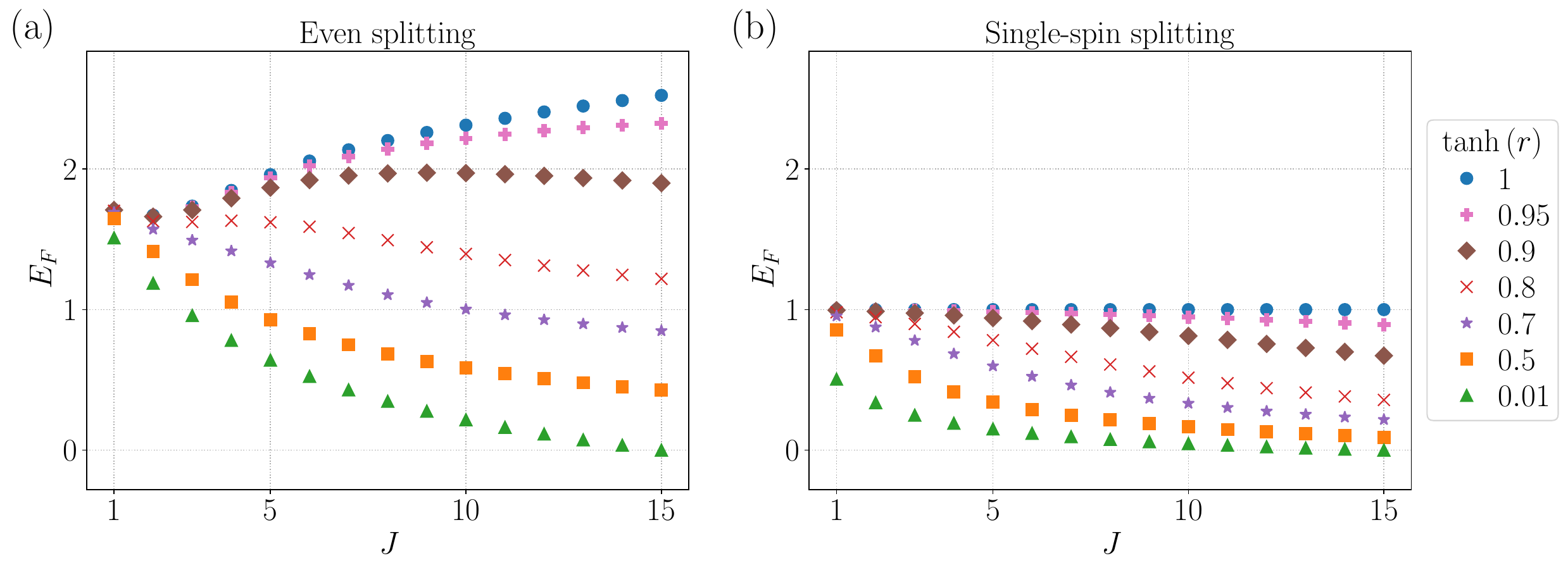}
    \caption{$E_F$ vs. $J$ for $N=30$ spin-squeezed states with (a) even splitting and (b) single-spin splitting. Here, $\tanh(r)$ indicates the degree of squeezing as defined in Eq.~(\ref{Dark state coefficients 1}).}
    \label{EJ SS states}
\end{figure*}

Another family of metrologically usefully entangled states is the spin-squeezed states. These states have one spin variance squeezed along a certain direction at the cost of increasing the spin variances along an orthogonal direction \cite{Metrology1}. They can be generated by a wide range of methods, such as applying coherent interaction \cite{CI1,CI2,CI3,CI4,CI5,CI6,CI7,CI8} or engineered dissipation \cite{CI3, Spin_squeezed_states_paper, Squeezed_light, SS1, CI6}. Since these methods do not require individual spin addressing, spin-squeezing has been an attractive approach to generate many-body entanglement and quantum sensing resources in unresolvable spin ensembles \cite{SS1, SS2, SS3}.

While the definition of a spin-squeezed state could depend on the generation process, here we consider the even-$N$ state that is stabilized by the collective spin dissipator \cite{Squeezed_light, Spin_squeezed_states_paper}:
\begin{equation}
    0=\mathcal{D}[\hat{\Sigma}[r]]\rho,
    \label{Master equation}
\end{equation}
where $r\geq 0$ determines the strength of squeezing, $\mathcal{D}[\hat{z}]\rho\equiv\hat{z}\rho\hat{z}^\dagger - \{\hat{z}^\dagger \hat{z},\rho\}/2$ is the Linblad superoperator, and $\hat{\Sigma}[r]\equiv\cosh(r)\hat{J}_- - \sinh(r)\hat{J}_+$ with $\hat{J}_{\pm}=\hat{J}_x\pm i\hat{J}_y$. Such states can be constructed by collectively coupling the spins to a squeezed reservoir \cite{Spin_squeezed_states_paper, CI6}.

For an ensemble with even $N$ initially prepared with a definite total angular momentum $J$, the stabilized PI state is given by Eqs.~(\ref{Superposition PI state}) and (\ref{Superposition PI state: components}) with the superposition amplitudes satisfying
\begin{equation}
    c^{(J)}_{-J+2k}(r)=\binom{J}{k}\sqrt{\binom{2J}{2k}^{-1}}\tanh^k(r)c^J_{-J}(r)
    \label{Dark state coefficients 1}
\end{equation}
for $k \in \{0,...,J\}$, and
\begin{equation}
    c^{(J)}_{-J+2k+1}(r)=0 \label{Dark state coefficients 2}
\end{equation}
for $k \in \{0,...,J-1\}$, while the coefficients obey the normalization $\sum_n |c_n|^2=1$. The state is a superposition of every other $M$ state. As $r$ increases, the state becomes more squeezed. We are interested in how the entanglement for the spin-squeezed states changes with $J$ and the degree of squeezing.

We show the results of a typical example in Fig.~\ref{EJ SS states}. For even splitting, the entanglement with low squeezing decreases monotonically with $J$. It is expected, because a low squeezed state is close to an angular momentum eigenstate with the lowest $J$, so the behavior resembles that in Fig.~\ref{EJ plot M=J}. However, as the degree of squeezing increases, the trend becomes more complicated. In the case of maximal squeezing, i.e., $\tanh(r)=1$, the entanglement initially drops but eventually increases towards maximum $J$.

This complicated behavior originates from an interplay of several effects. First, as squeezing increases, magnetization eigenstates with $M$ other than $-J$ will populate. As discussed in Sec.~\ref{Entanglement vs. total angular momentum}, entanglement of some $M$ states can be non-monotonic in $J$. Besides, the number of magnetization eigenstates involved in the squeezed-state superposition scales linearly as $J$, i.e., $M = \{-J, -J+2, ..., J\}$. When squeezing is high, each of these components will have a significant superposition amplitude, so as $J$ increases, there will be more superposition components and thereby greater extrinsic entanglement.

For single-spin splitting, entanglement generally decreases with $J$. While entanglement does increase with the degree of squeezing, the competition effect observed in even splitting does not occur. This is because the number of Schmidt components is always limited to two, so increasing $J$ does not contribute additional superposition. However, in the case of maximal squeezing, the entanglement is maximal for all (integer) $J$. We prove this in Appendix \ref{Appendix:dark} with the explicit expression of the maximal squeezing states.

\section{\label{Section: conclusion}Conclusions}

Quantum entanglement is both an essential resource in quantum information processing and an intricate fundamental phenomenon. One important task is to quantify the amount of entanglement in a many-body quantum state. Unfortunately, realistic quantum states are often mixed, and quantifying mixed-state entanglement is extremely challenging. In this work, we provide an exact quantification for a huge class of PI states. Such states are fundamentally mixed but practically relevant; they are found in realistic spin-ensemble quantum technological platforms where resolving individual spins is infeasible. By taking advantage of the structure of the PI states, we show that the entanglement of formation of those whose collective properties are known is computable. The amount is exactly equal to the average entanglement entropy of its standard decomposition.

By using the exact formalism, we study the relationship between entanglement and different attributes of the system, and find surprising behaviors in various scenarios. Generally, we show that the entanglement of generic PI states behaves quite differently from the well-studied Dicke states. Specifically, we find that PI state entanglement can exhibit zigzag patterns as the parity of the particle number or the partition number changes (Figs.~\ref{EN min M plot} and \ref{En plot}), and that its behavior for low $J$ can alter significantly depending on the value of $M$ and the partition number (Fig.~\ref{EJ plot M=J}b).

Beyond the test cases conducted in this work, we envision that our techniques can find more applications in, for example, understanding the role of entanglement in ensemble-based quantum technologies, tracking information flow in collective spin dynamics, and benchmarking entanglement witnesses in spin ensembles.

\begin{acknowledgments}
This work is supported by the Natural Sciences and Engineering Research Council of Canada (Grant No. NSERC RGPIN-2021-02637) and Canada Research Chairs (Grant No. CRC-2020-00134).
\end{acknowledgments}

\appendix
\section{\label{Appendix:rise}Addition of angular momenta and the arising of permutational symmetry}
\begin{figure}[htp!]
    \centering
    \includegraphics[width=7.5cm]{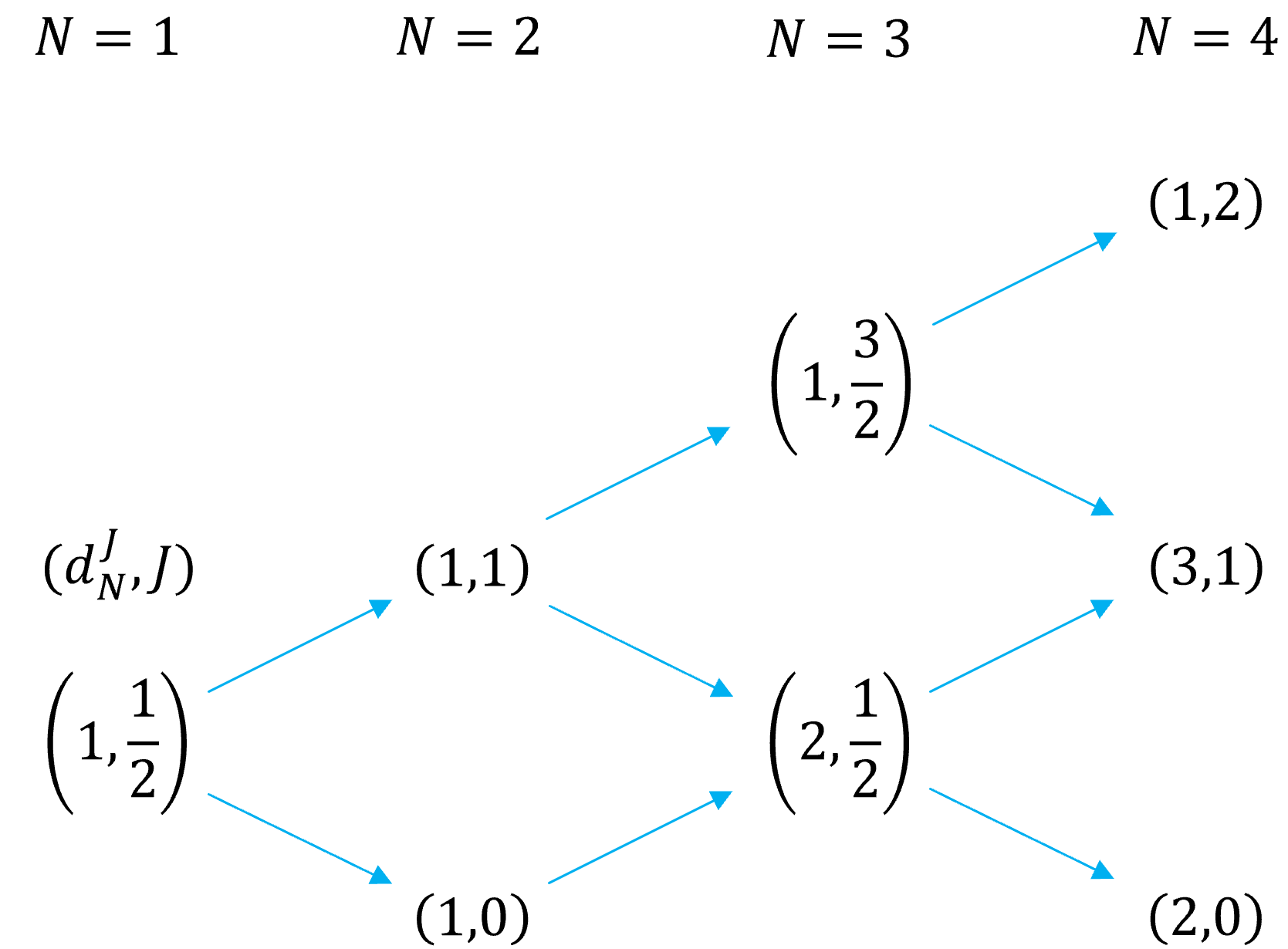}
    \caption{Diagram showing the degeneracy structure $(d^J_N, J)$ of a spin-1/2 ensemble as particles are added sequentially. Each node corresponds to a possible total angular momentum $J$ for a given particle number $N$, and blue arrows indicate the possible change of total angular momentum $(\pm 1/2)$ when adding a spin-1/2 particle. The number of ways to combine $N$ spin-1/2 particles to form a total angular momentum $J$ is equal to the number of all distinct paths $d^J_N$ from the starting node at $N=1$ to the specified node.}
    \label{degeneracy structure}
\end{figure}

According to the rules of angular momentum addition, the Hilbert space of $N$ spin-1/2 particles can be decomposed as follows \cite{PI_definition2,Duality1,Duality2}:
\begin{equation}
    \mathcal{H}=(\mathbb{C}^2)^{\otimes N}= \bigoplus^{N/2}_{J=J_{\text{min}}} \mathcal{H}_J \otimes \mathcal{K}_J,
    \label{N spin isomorphism}
\end{equation}
where $\mathcal{H}_J$ is the irreducible SU(2) representation of spin $J$, and the multiplicative space $\mathcal{K}_J$, where the irreducible representations of the symmetric group $S_N$ act, encodes the degeneracy associated with different coupling paths. The dimension of $\mathcal{H}_J$ is $2J+1$, equal to the number of allowed values of $M\in\{-J,-J+1,...,J-1,J\}$. The dimension of the multiplicative space is
\begin{equation}
    \text{dim}(\mathcal{K}_J)=d^J_N=\frac{N!(2J+1)}{\left(\frac{N}{2}-J\right)!(\frac{N}{2}+J+1)!}, \label{degeneracy}
\end{equation}
which means that there are $d^J_N$ ways to combine $N$ spin-1/2 particles to form a total spin-$J$ state. This structure of angular momentum addition and its associated degeneracies is illustrated in Fig.~\ref{degeneracy structure}. We label distinct paths that lead to the same total spin state by $i\in\{1,...,d^J_N\}$.

For example, when adding spin-1/2 particles, their spins can be either symmetrically aligned to form a spin-1 state or anti-symmetrically to form a spin-0 state. This can be compactly expressed as $\frac{1}{2}\otimes \frac{1}{2}=1\oplus 0$. For three spin-1/2 particles, we can first combine two into either an effective spin-1 or spin-0, and then combine the third spin. In the case of spin-1, the total spin can either be 3/2 or 1/2, while the spin-0 case can only lead to a total spin-1/2. This structure is expressed as $\frac{1}{2}\otimes \frac{1}{2}\otimes \frac{1}{2}=(1\oplus 0)\otimes \frac{1}{2}=(\frac{3}{2}\oplus \frac{1}{2})\oplus \frac{1}{2}$. We see that two orthogonal subspaces share the same total spin as 1/2. In Fig.~\ref{degeneracy structure}, these two distinct subspaces correspond to different angular momentum addition paths. We label them by a degeneracy index $i$, e.g, $i=1$ for the spin-1 path and $i=2$ for the spin-0 path.

When individual spins cannot be addressed and the degeneracy index $i$ is experimentally inaccessible, one has no information about the path of angular momentum addition that the total spin state takes. As a result, the ensemble has to be described by an equal-weight mixed state over all $i$. This loss of path information leads to permutational symmetry because, intuitively, neither the total angular momentum $J$ nor the degeneracy distribution over $i$ changes under particle permutations.

This permutational symmetry is captured by Eq.~(\ref{N spin isomorphism}). It shows that the total Hilbert space is spanned by the basis $\{\ket{J,M,N}\otimes\ket{i} = \ket{J,M,N,i}\}$, and the most general PI states should take the block-diagonal form \cite{PI_definition1,PI_definition2,Duality1}:
\begin{equation}
    \rho_{\text{PI}}=\bigoplus^{N/2}_{J=J_{\text{min}}}p_J\left(\rho_J\otimes \mathbbm{1}_{\mathcal{K}_J}\right), \label{General PI state}
\end{equation}
where $p_J$ is the probability of finding the ensemble in the $J$ subspace. For each $J$, we have a block-diagonal density matrix consisting of $d^J_N$ subblocks for each degeneracy index $i$:
\begin{equation}
\begin{split}
    \rho_{\text{PI}}^J &\equiv \rho_J\otimes\mathbbm{1}_{\mathcal{K}_J}\\
    &=\frac{1}{d^J_N}\sum_{M,M'=-J}^{J}\sum_{i=1}^{d^J_N}\rho_{M,M'}^J\ket{J,M,N,i}\bra{J,M',N,i}.
    \label{General PI state of J}
\end{split}
\end{equation}
We show the matrix representation of $\rho^J_{\text{PI}}$ in Fig.~\ref{Block diagonal form of PI state}. It can be readily checked that Eqs.~(\ref{General PI state}) and (\ref{General PI state of J}) both satisfy $\rho=\hat{P}_\pi \rho \hat{P}_\pi^\dagger$ for all particle permutations $\pi \in S_N$, so they are indeed permutationally invariant. Note that the density matrix elements $\rho_{M,M'}^J$ are diagonal in $J$, which means that PI states do not have any coherence between different $J$.

\begin{figure}[htp!]
    \centering
    \[ 
    \setlength\arraycolsep{0pt}
    \rho_{\text{PI}}^J = 
    \begin{pmatrix}
    \SquareBox[draw=black, thick, fill=customblue!80]{~~~$\rho_{i=1}^J$~~~} &  &  &  \\
     & ~\SquareBox[draw=black, thick, fill=customblue!80]{~~~$\rho_{i=2}^J$~~~} &  &  \\
    &  & ~~~\ddots~~ &  \\
    &  &  & \SquareBox[draw=black, thick, fill=customblue!80]{~~$\rho_{i=d^J_N}^J$~~}
    \end{pmatrix}
    \]
    \caption{Block-diagonal form of $\rho_{\text{PI}}^J$ in the coupled spin basis. For each $i$, there is an associated subblock of dimension $(2J+1)\times(2J+1)$. The blank entries are filled with zeros.}
    \label{Block diagonal form of PI state}
\end{figure}

\section{\label{Appendix:derivation}Derivation of Eq.~(\ref{Gold equation for entanglement: Superposition})}
In Eq.~(\ref{Exact entanglement}), we see that the entanglement of formation for PI states is equal to the average entanglement entropy. This means that we can evaluate the entanglement of formation by averaging the entanglement entropy over all the $i$ components, each occurring with equal probability $1/d^{J}_N$ (cf.~Eq.~(\ref{Superposition PI state})):
\begin{equation}
    E_F(\rho^J_{\text{PI}};n)=\frac{1}{d^J_N}\sum_{i=1}^{d^J_N} E\left(\ket{\psi^J_i};n\right). \label{Derivation:unmassaged}
\end{equation}
We know from Eq.~(\ref{Basis state being split with i}) that changing the splitting changes how the degenerate basis states are expressed, but here how $i$ and $n$ are related is unclear. To make the dependence on the splitting explicit, instead of enumerating all $i$, we can equivalently enumerate all possible $(j_1, j_2)$ decompositions and their associated degeneracies $(i^{j_1},i^{j_2})$, where $i^{j_1}=1,2,...,d^{j_1}_{n}$ and $i^{j_2}=1,2,...,d^{j_2}_{N-n}$. That is, rather than counting the global degeneracy index, we count the local degeneracy indices, rewriting the sum $\sum_{i=1}^{d^J_N} E\left(\ket{\psi^J_i};n\right)$ as
\begin{equation}
    \sum_{j_1=j_{1,\text{min}}}^{n/2}\sum_{j_2=j_{2,\text{min}}}^{(N-n)/2} \sum_{i^{j_1}=1}^{d^{j_1}_n} \sum_{i^{j_2}=1}^{d^{j_2}_{N-n}} E\left(\ket{\psi^{(j_1,j_2)}_{(i^{j_1},i^{j_2})}}\right). \label{Derivation:full}
\end{equation}
This equivalence can be readily checked, since $\sum_{j_1}^{n/2}\sum_{j_2}^{(N-n)/2}d^{j_1}_{n}d^{j_2}_{N-n}=d^J_N$ (summing only over $(j_1,j_2)$ satisfying Rule II), so the number of combinations matches.

The next step is to realize that we do not need to sum over all possible $(i^{j_1},i^{j_2})$ because the CG coefficients $\leftindex^{J,M}C_{j_1,m_1;j_2,m_2}$ are only determined by $(j_1,j_2)$. This means that, for a given $(j_1,j_2)$, all subensemble degeneracy indices give the same entanglement. Since for every $(j_1,j_2)$ there are $d^{j_1}_{n}d^{j_2}_{N-n}$ degeneracies, Eq.~(\ref{Derivation:full}) can be reduced to
\begin{equation}
    \sum_{j_1=j_{1,\text{min}}}^{n/2}\sum_{j_2=j_{2,\text{min}}}^{(N-n)/2}d^{j_1}_n d^{j_2}_{N-n} E\left(\ket{\psi^{(j_1,j_2)}}\right), \label{Derivation:simplify}
\end{equation}
which is significantly simpler to compute, as it avoids counting exponentially many $(i^{j_1},i^{j_2})$. Substituting this expression back into Eq.~(\ref{Derivation:unmassaged}), the entanglement of formation takes the form
\begin{equation}
    E_F(\rho^J_{\text{PI}};n)=\sum_{(j_1,j_2)}\frac{d^{j_1}_{n}d^{j_2}_{N-n}}{d^J_N}E\left(\ket{\psi^{(j_1,j_2)}}\right). \label{Derivation:reminiscence}
\end{equation}
Here, the weights $d^{j_1}_{n}d^{j_2}_{N-n}/{d^J_N}$ are the probabilities that the two subensembles are found to have angular momenta $j_1$, $j_2$.

Lastly, we determine the entanglement entropy $E\left(\ket{\psi^{(j_1,j_2)}}\right)$. We first insert the decomposition in Eq.~(\ref{Basis state being split with i}) into Eq.~(\ref{Superposition PI state: components}) to obtain
\begin{align}
    \ket{\psi^{J}_i}
    &=\sum_{M=-J}^{J}c_M\sum_{m_1=-j_1}^{j_1}\sum_{m_2=-j_2}^{j_2}\leftindex^{J,M}C_{j_1,m_1;j_2,m_2} \nonumber \\
    &\quad\times\ket{j_1,m_1,n,i^{j_1}}\otimes \ket{j_2,m_2,N-n,i^{j_2}}. \label{Derivation:state}
\end{align}
Then, we drop the degeneracy indices $i$, $i^{j_1}$, $i^{j_2}$ and focus on the subensemble angular momenta, rewriting Eq.~(\ref{Derivation:state}) as
\begin{align}
    \ket{\psi^{(j_1,j_2)}}
    &=\sum_{M=-J}^{J}c_M\sum_{m_1=-j_1}^{j_1}\sum_{m_2=-j_2}^{j_2}\leftindex^{J,M}C_{j_1,m_1;j_2,m_2} \nonumber \\
    &\quad\times\ket{j_1,m_1,n}\otimes \ket{j_2,m_2,N-n}. \label{Derivation:state_no_i}
\end{align}
According to Eq.~(\ref{Von Neumann entropy of entanglement}), this state's entanglement entropy $E\left(\ket{\psi^{(j_1,j_2)}}\right)=-\sum_k \nu_{k}^{(j_1,j_2)} \log_2{\nu_{k}^{(j_1,j_2)}}$ can be calculated by finding the eigenvalues $\nu_k^{(j_1,j_2)}$ of the reduced density matrix $\sigma_1^{(j_1,j_2)}\equiv \Tr_2{\{\ketbra{\psi^{(j_1,j_2)}}\}}$. Replacing $\rho^J_{M,M'}\equiv c_M c_{M'}^*$ and using $m_2=M-m_1$ (Rule I), we obtain Eq.~(\ref{reduced density matrix for superposition}). Finally, substituting the entanglement entropy into Eq.~(\ref{Derivation:reminiscence}), we derive Eq.~(\ref{Gold equation for entanglement: Superposition}).

\section{\label{Appendix:single}Formula for entanglement of formation with single-spin splitting}
For single-spin splitting, there are at most two possible pairs of $(j_1,j_2)$ that satisfy Rule I, and we can write down a general analytical expression for the entanglement of formation.

According to Rule I, for $J=0$, only $j_2 = 1/2$ is possible; for $J=N/2$, only $j_2 = (N-1)/2$ is possible. For other intermediate $J$, there are two possible values for $j_2$: either $j_2=J+1/2$ or $j_2=J-1/2$. The probabilities of measuring the two outcomes are
\begin{equation}
    j_2=J+\frac{1}{2}:\quad\frac{d^{J+1/2}_{N-1}}{d^J_N}=\frac{1}{N}\left(\frac{2J+2}{2J+1}\right)\left(\frac{N}{2}-J\right).
    \label{j2 first possibility with n=1}
\end{equation}
\begin{equation}
    j_2=J-\frac{1}{2}:\quad\frac{d^{J-1/2}_{N-1}}{d^J_N}=\frac{1}{N}\left(\frac{2J}{2J+1}\right)\left(\frac{N}{2}+J+1\right),
    \label{j2 second possibility with n=1}
\end{equation}
According to Eq.~(\ref{Gold equation for entanglement}), the entanglement of formation of an angular momentum eigenstate can be calculated by the average entanglement entropy of the two possibilities; this will be a function of the CG coefficients:
\begin{widetext}
    \begin{align}
\begin{split}
    E_F(\rho_{\text{PI}}^{J,M};n=1)=&-\frac{1}{N(2J+1)}\left\{(2J+2)\left(\frac{N}{2}-J\right)\left[\left( ^{J,M}C_{\frac{1}{2},\frac{1}{2}\,;J+\frac{1}{2},M-\frac{1}{2}}\right)^2\log_2{\left( ^{J,M}C_{\frac{1}{2},\frac{1}{2}\,;J+\frac{1}{2},M-\frac{1}{2}}\right)^2}\right.\right.\\
    &\qquad\qquad\qquad\qquad+\left.\left( ^{J,M}C_{\frac{1}{2},-\frac{1}{2}\,;J+\frac{1}{2},M+\frac{1}{2}}\right)^2\log_2{\left( ^{J,M}C_{\frac{1}{2},-\frac{1}{2}\,;J+\frac{1}{2},M+\frac{1}{2}}\right)^2}\right]\\
    &\qquad\qquad\qquad+2J\left(\frac{N}{2}+J+1\right)\left[\left( ^{J,M}C_{\frac{1}{2},\frac{1}{2}\,;J-\frac{1}{2},M-\frac{1}{2}}\right)^2\log_2{\left( ^{J,M}C_{\frac{1}{2},\frac{1}{2}\,;J-\frac{1}{2},M-\frac{1}{2}}\right)^2}\right.\\
    &\qquad\qquad\qquad\qquad+\left.\left( ^{J,M}C_{\frac{1}{2},-\frac{1}{2}\,;J-\frac{1}{2},M+\frac{1}{2}}\right)^2\log_2{\left( ^{J,M}C_{\frac{1}{2},-\frac{1}{2}\,;J-\frac{1}{2},M+\frac{1}{2}}\right)^2}\right]\biggl\}\\
    =&-\frac{1}{N(2J+1)}\left\{\left(\frac{N}{2}-J\right)\left[\left(J-M+1\right)\log_2{\left(\frac{J-M+1}{2J+2}\right)}\right.\right.+\left.\left(J+M+1\right)\log_2{\left(\frac{J+M+1}{2J+2}\right)}\right]\\
    &\quad\quad+\left(\frac{N}{2}+J+1\right)\left[\left( J+M\right)\log_2{\left(\frac{J+M}{2J}\right)}\right.+\left.\left(J-M\right)\log_2{\left( \frac{J-M}{2J}\right)}\right]\biggl\}.
    \label{entanglement n=1}
\end{split}
\end{align}
\end{widetext}
In the second line, we have used
\begin{equation}
    \left( ^{J,M}C_{\frac{1}{2},\frac{1}{2}\,;J+\frac{1}{2},M-\frac{1}{2}}\right)^2=\frac{J-M+1}{2J+2},
\end{equation}
\begin{equation}
    \left( ^{J,M}C_{\frac{1}{2},-\frac{1}{2}\,;J+\frac{1}{2},M+\frac{1}{2}}\right)^2=\frac{J+M+1}{2J+2},
\end{equation}
\begin{equation}
    \left( ^{J,M}C_{\frac{1}{2},\frac{1}{2}\,;J-\frac{1}{2},M-\frac{1}{2}}\right)^2=\frac{J+M}{2J},
\end{equation}
and
\begin{equation}
    \left( ^{J,M}C_{\frac{1}{2},-\frac{1}{2}\,;J-\frac{1}{2},M+\frac{1}{2}}\right)^2=\frac{J-M}{2J}.
\end{equation}

Equation (\ref{entanglement n=1}) allows any set of parameters $\{J,M,N,n=1\}$. It underlies all the surprising phenomena that are specific to single-spin splitting.

\section{\label{Appendix:superposition}Extrinsic and intrinsic entanglement}
In Sec. \ref{Section: Superposition states}, we mentioned that if each subensemble component $\ket{j_1,m_1}$ and $\ket{j_2,m_2}$ in Eq.~(\ref{Superposition state discussion}) is associated with only one $M$ state, i.e., if $\ket{\psi_i^J}$ is in Schmidt form, then the entanglement entropy is the sum of the extrinsic and intrinsic entanglement entropies. Here, we quickly prove this statement.

Since Eq.~(\ref{Superposition state discussion}) is already in Schmidt form, the Schmidt coefficients are simply $\left|c_M \leftindex^{J,M}C_{j_1,m_1;j_2,m_2}\right|^2$ for each $M$. A direct calculation of the entanglement entropy gives
\begin{widetext}
    \begin{align}
    E\left(\ket{\psi^J_i}\right)=&-\sum_{M}\sum_{m_1,m_2}\left|c_M \leftindex^{J,M}C_{j_1,m_1;j_2,m_2}\right|^2 \log_2{\left|c_M \leftindex^{J,M}C_{j_1,m_1;j_2,m_2}\right|^2} \nonumber \\
    =&-\sum_M \left|c_M\right|^2\sum_{m_1,m_2}\left(\leftindex^{J,M}C_{j_1,m_1;j_2,m_2}\right)^2 \log_2{\left|c_M\right|^2}+\left(\leftindex^{J,M}C_{j_1,m_1;j_2,m_2}\right)^2\log_2{\left(\leftindex^{J,M}C_{j_1,m_1;j_2,m_2}\right)^2} \nonumber\\
    =&-\sum_M \left|c_M\right|^2 \log_2{\left|c_M\right|^2}-\sum_M \left|c_M\right|^2\sum_{m_1,m_2}\left(\leftindex^{J,M}C_{j_1,m_1;j_2,m_2}\right)^2\log_2{\left(\leftindex^{J,M}C_{j_1,m_1;j_2,m_2}\right)^2}.
    \end{align}
\end{widetext}
From the second to the third line, we applied the identity $\sum_{m_1,m_2}\left(\leftindex^{J,M}C_{j_1,m_1;j_2,m_2}\right)^2=1$. The first term is the extrinsic entanglement entropy due to the superposition amplitudes $c_M$, and the second term is the weighted intrinsic entanglement entropy of each $M$ state.

\section{\label{Appendix:dark}Squeezed state with single-spin splitting}
For single-spin splitting, the maximally squeezed state has the following form (here we show the $j_2=J+1/2$ case as an example):
\begin{widetext}
    \begin{align}
        &\ket{\psi[J,N,i=1;r=1]} \nonumber\\
        =&c^{(J)}_{-J}(1)\leftindex^{J,-J}C_{\frac{1}{2},\frac{1}{2};J+\frac{1}{2},-J-\frac{1}{2}}\ket{\uparrow}\ket{J+\frac{1}{2},-J-\frac{1}{2}}+\dots+c^{(J)}_J(1)\leftindex^{J,J}C_{\frac{1}{2},\frac{1}{2};J+\frac{1}{2},J-\frac{1}{2}}\ket{\uparrow}\ket{J+\frac{1}{2},J-\frac{1}{2}} \nonumber\\
        +&c^{(J)}_{-J}(1)\leftindex^{J,-J}C_{\frac{1}{2},-\frac{1}{2};J+\frac{1}{2},-J+\frac{1}{2}}\ket{\downarrow}\ket{J+\frac{1}{2},-J+\frac{1}{2}}+\dots+c^{(J)}_{J}(1)\leftindex^{J,J}C_{\frac{1}{2},-\frac{1}{2};J+\frac{1}{2},J+\frac{1}{2}}\ket{\downarrow}\ket{J+\frac{1}{2},J+\frac{1}{2}} \nonumber\\[5pt]
        =&\frac{1}{\sqrt{2}}\ket{\uparrow}\left(\sqrt{2}\,c^{(J)}_{-J}(1)\leftindex^{J,-J}C_{\frac{1}{2},\frac{1}{2};J+\frac{1}{2},-J-\frac{1}{2}}\ket{J+\frac{1}{2},-J-\frac{1}{2}}+\dots+\sqrt{2}\,c^{(J)}_J(1)\leftindex^{J,J}C_{\frac{1}{2},\frac{1}{2};J+\frac{1}{2},J-\frac{1}{2}}\ket{J+\frac{1}{2},J-\frac{1}{2}}\right) \nonumber\\
        +&\frac{1}{\sqrt{2}}\ket{\downarrow}\left(\sqrt{2}\,c^{(J)}_{-J}(1)\leftindex^{J,-J}C_{\frac{1}{2},-\frac{1}{2};J+\frac{1}{2},-J+\frac{1}{2}}\ket{J+\frac{1}{2},-J+\frac{1}{2}}+\dots+\sqrt{2}\,c^{(J)}_{J}(1)\leftindex^{J,J}C_{\frac{1}{2},-\frac{1}{2};J+\frac{1}{2},J+\frac{1}{2}}\ket{J+\frac{1}{2},J+\frac{1}{2}}\right).
        \label{dark state s=1}
    \end{align}
\end{widetext}
Using the identity $c^{(J)}_M(1)=c^{(J)}_{-M}(1)$, the normalization condition $\sum_M\left| c^{({J})}_M \right|^2=1$, the orthonormality of CG coefficients $\left|\leftindex^{J,M}C_{\frac{1}{2},\pm\frac{1}{2};J+\frac{1}{2},M\mp\frac{1}{2}}\right|^2+\left|\leftindex^{J,-M}C_{\frac{1}{2},\pm\frac{1}{2};J+\frac{1}{2},-M\mp\frac{1}{2}}\right|^2=1$, and $\left|\leftindex^{J,0}C_{\frac{1}{2},\pm\frac{1}{2};J+\frac{1}{2},\mp\frac{1}{2}}\right|^2=1/2$ for the even $J$ case, it can be readily checked that the subensemble 2 components inside the two brackets are properly normalized. For any integer $J$, this state has two Schmidt coefficients 1/2, so it is maximally entangled. The $j_2=J-1/2$ case has a similar form and is also maximally entangled for any integer $J$. Therefore, the average entanglement of the two is maximum, as we have seen in Fig.~\ref{EJ SS states}b.


\bibliography{apssamp}

\end{document}